\documentclass[review,3p]{elsarticle}

\usepackage{multirow}
\usepackage{caption}

\journal{Elsevier}

\usepackage{color}

\begin{document}

\begin{frontmatter}

\title{Bridging 2D and 3D Segmentation Networks for Computation-Efficient Volumetric Medical Image Segmentation: An Empirical Study of 2.5D Solutions}

\author[mymainaddress]{Yichi Zhang}

\author[mymainaddress]{Qingcheng Liao}

\author[mymainaddress]{Le Ding}

\author[mymainaddress,mysecondaryaddress,mythirdaddress,myfourthaddress]{Jicong Zhang\corref{mycorrespondingauthor}}
\cortext[mycorrespondingauthor]{Corresponding author at: School of Biological Science and Medical Engineering, Beihang University, Beijing, China. \\ E-mail addresses: \textit{coda1998@buaa.edu.cn} (Yichi Zhang), \textit{jicongzhang@buaa.edu.cn} (Jicong Zhang) }

\address[mymainaddress]{School of Biological Science and Medical Engineering, Beihang University, Beijing, China}
\address[mysecondaryaddress]{Hefei Innovation Research Institute, Beihang University, Hefei, China}
\address[mythirdaddress]{Beijing Advanced Innovation Centre for Biomedical Engineering, Beijing, China}
\address[myfourthaddress]{Beijing Advanced Innovation Centre for Big Data-Based Precision Medicine, Beijing, China}

\begin{abstract}
Recently, deep convolutional neural networks have achieved great success for medical image segmentation. However, unlike segmentation of natural images, most medical images such as MRI and CT are volumetric data. In order to make full use of volumetric information, 3D CNNs are widely used. However, 3D CNNs suffer from higher inference time and computation cost, which hinders their further clinical applications. Additionally, with the increased number of parameters, the risk of overfitting is higher, especially for medical images where data and annotations are expensive to acquire. To issue this problem, many 2.5D segmentation methods have been proposed to make use of volumetric spatial information with less computation cost. Despite these works lead to improvements on a variety of segmentation tasks, to the best of our knowledge, there has not previously been a large-scale empirical comparison of these methods. In this paper, we aim to present a review of the latest developments of 2.5D methods for volumetric medical image segmentation. Additionally, to compare the performance and effectiveness of these methods, we provide an empirical study of these methods on three representative segmentation tasks involving different modalities and targets. Our experimental results highlight that 3D CNNs may not always be the best choice. Despite all these 2.5D methods can bring performance gains to 2D baseline, not all the methods hold the benefits on different datasets. We hope the results and conclusions of our study will prove useful for the community on exploring and developing efficient volumetric medical image segmentation methods.
\end{abstract}


\end{frontmatter}

\section{Introduction}

Medical image segmentation aims to extract meaningful objects or regions from the image, which is a fundamental and important step for many clinical applications like computer-assisted diagnosis, treatment planning and surgery navigation.
For example, accurate location and segmentation of abdominal anatomy from CT images is helpful in cancer diagnosis and treatment \cite{ wolz2013automated,ma2021abdomenct}. Atrial segmentation plays an important role in the medical management of patients \cite{bernard2018deep,lalande2021deep}. 
However, manual segmentation of medical images is a labor-intensive and tedious task, and subjects to variations of observers. This has motivated many researches on automatic segmentation methods. 
In the past several years, convolutional neural networks (CNNs)$\footnote{In order to unify the statement, in this paper, CNNs refer specifically to convolution neural networks for medical image segmentation. \\ }$ have been proposed and achieved state-of-the-art performance in many medical image segmentation tasks \cite{bernard2018deep,lalande2021deep,wang2019benchmark,heller2021state}.
Most of these segmentation methods are based on U-Net \cite{ronneberger2015u-net} or its variants, using encoder-decoder architecture and incorporating multi-scale features by skip connections, which is proven to be a strong baseline network for medical image segmentation \cite{isensee2020nnu}. 

\begin{figure}
\includegraphics[width=16cm]{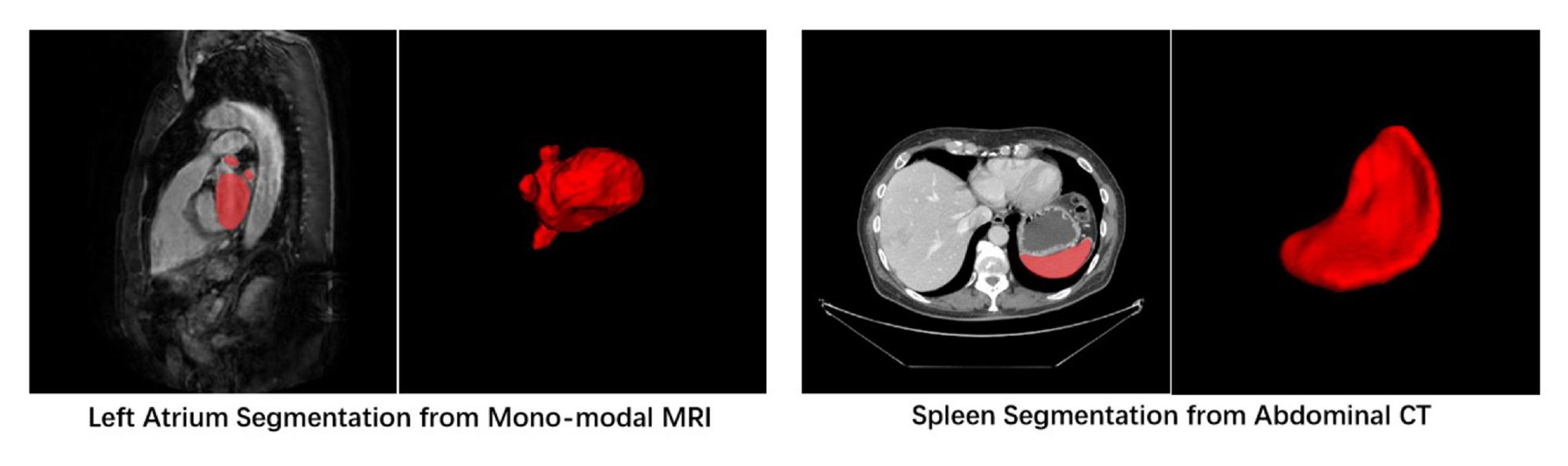}
\centering
\captionsetup{font={small}}
\caption{Examples for volumetric medical image segmentation. The image with 2D label denoted by red and corresponding 3D reconstruction of the label are shown in the figure.}
\label{Fig1}
\end{figure}

With the development of medical imaging technology, 3D medical images such as Computed Tomography (CT) and Magnetic Resonance Imaging (MRI) have been widely used in clinical approaches. 
Figure \ref{Fig1} presents two different examples of volumetric medical image segmentation with corresponding 3D reconstruction of segmentation results.
Automatic segmentation of volumetric medical images has become increasingly important for biomedical applications \cite{hesamian2019deep,tajbakhsh2020embracing}. To deal with volumetric input, two main strategies are applied. The first method is cutting the 3D volume into 2D slices and training 2D CNNs for segmentation based on intra-slice information. Another strategy is to extend the network structure to volumetric data by using 3D convolutions and train 3D CNNs for segmentation based on volumetric images \cite{iek20163D,milletari2016v}.
Both of these two methods have their own advantages and disadvantages. 2D CNNs have lighter computation and higher inference speed. However, the information between adjacent slices is neglected, which hinders the improvement of segmentation accuracy. Besides, 2D segmentation results are prone to discontinuity in 3D space \cite{qu2021surgical}.
Although 3D CNNs have the perception of volumetric spacial information, there are still some shortcomings. Due to the increased dimension, 3D CNNs require higher computation cost, resulting in higher inference latency \cite{yu2019thickened,zhang2020saunet}. Besides, the large number of parameters may result in higher risk of overfitting, especially when encountering small datasets. Moreover, the GPU requirement of 3D CNNs is impractically expensive, which hinders their further clinical application.

To bridge the gap between 2D and 3D CNNs, many 2.5D segmentation methods (also known as pseudo 3D methods) are proposed for efficient volumetric medical image segmentation by designing new architecture or using strategies to fuse volumetric information into 2D CNNs. In this way, the models could enjoy both light computation cost and time consumption of 2D CNNs and the ability to capture spacial contextual information of 3D CNNs. 
However, these methods are tested on different datasets for different segmentation tasks. Their performance in other medical image segmentation tasks is unknown. Besides, the comparison experiments of most 2.5D methods are only compared with 2D and 3D CNNs, the comparison between these methods have not been verified. Therefore, in practical applications, we do not know which method should be selected to obtain better performance gains. In addition, although 2.5D CNNs obviously suffer from less time consumption compared with 3D CNNs, the improvement has not been quantitatively studied. This has hindered the further research and application of these methods.

In this paper, our contributions are summarized as follows:

\begin{itemize}
\item We summarize the latest developments about 2.5D methods for volumetric medical image segmentation and divide these methods into three categories: multi-view fusion, incorporating inter-slice information and fusing 2D/3D features.

\item We make a large-scale empirical comparison of these 2.5D methods with 2D and 3D CNNs on 3 representative public datasets involving different modalities (CT and MRI) and targets (cardiac, prostate and abdomen) to systematically evaluate the performance of these methods on different medical image segmentation tasks.

\end{itemize}

\section{Related Work}

For segmentation of volumetric medical images, 3D CNNs could capture richer contextual information and improve the performance over the 2D CNNs, while higher memory requirements and time consumption are required \cite{xia2018bridging}. To issue this problem, many studies have suggested the usage of 2.5D segmentation methods that fuse volumetric spatial information into 2D CNNs to improve the accuracy while reducing the computational cost.
In this section, we present an overview of 2.5D segmentation methods based on the criteria that the method should be general and can be applied to different 3D segmentation tasks. Several related methods also use 2.5D methods for specific task like blood vessel segmentation \cite{angermann2019projection}. The evaluation of these tailored methods is beyond the scope of this paper.
In the following subsections, we give a review of 2.5D segmentation methods that will be evaluated in the following experiments. We divided these methods into three categories: multi-view fusion, incorporating inter-slice information and fusing 2D/3D features.

\subsection{Multi-view fusion}

\begin{figure}
\includegraphics[width=16cm]{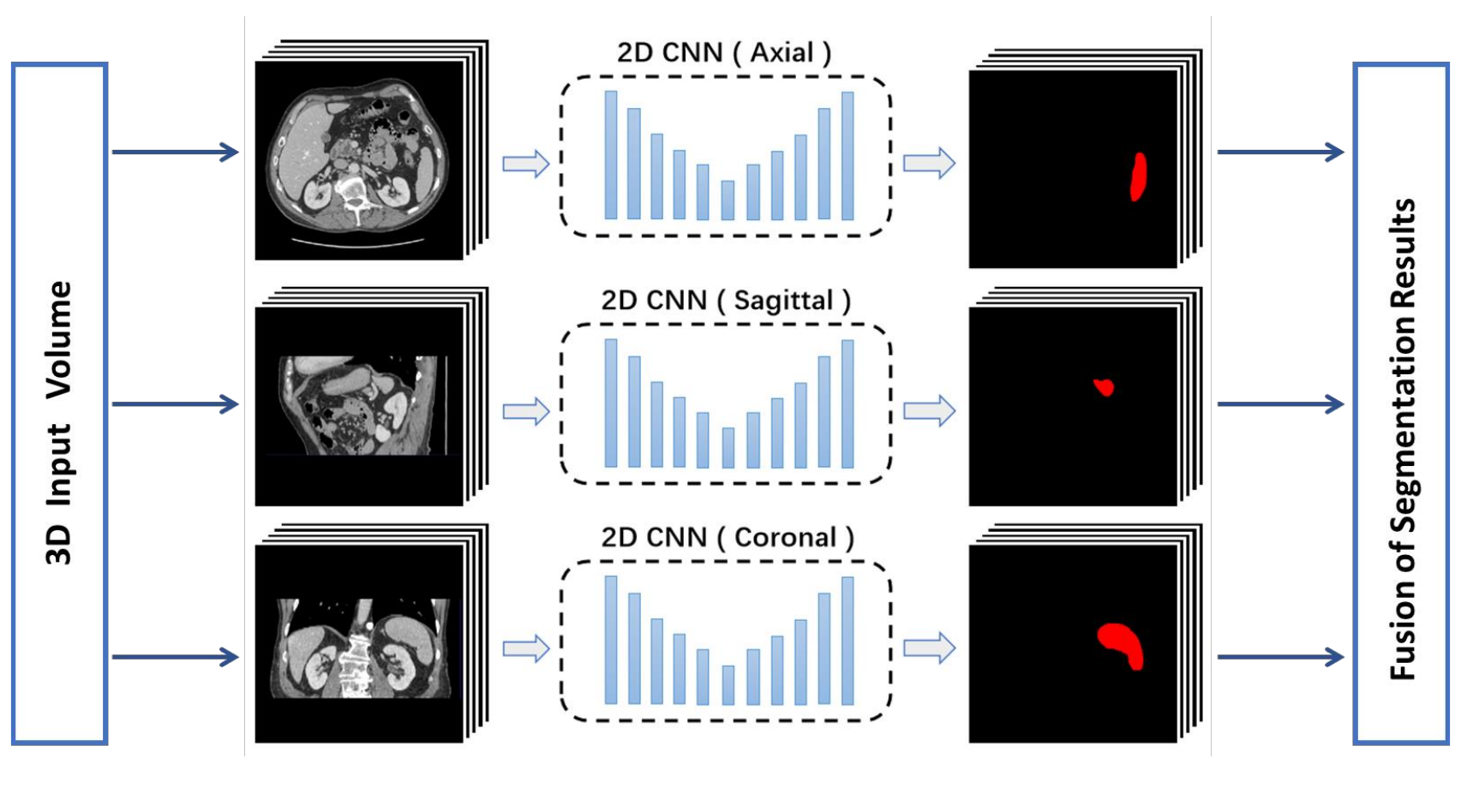}
\centering
\captionsetup{font={small}}
\caption{The overall workflow of multi-view fusion. The 3D input volume is divided into 2D slices from axial, sagittal and coronal planes to train the corresponding 2D CNNs. Then the segmentation results from different views are ensembled to get the final 2.5D segmentation result.}
\label{Fig2}
\end{figure}

To incorporate volumetric spatial information into 2D CNNs, a simple and intuitive solution is multi-view fusion (MVF). By generating slice-by-slice predictions based on 2D CNNs from different image views, and then fusing the segmentation results by means of majority voting, the final 2.5D segmentation result is obtained. Generally, the strategy is performed on three orthogonal planes, including the sagittal (S), coronal (C) and axial (A) planes. For the segmentation of each plane, the information of two axes could be used. Therefore, by combining the results of sagittal slices, coronal slices and axial slices together, the spatial information of 3D volumes can be fully utilized, so as to obtain better segmentation results compared with 2D CNNs. The overall workflow of multi-view fusion is shown in Figure \ref{Fig2}.

Multi-view fusion of three orthogonal planes has been applied in many 3D medical image segmentation tasks \cite{xia2018bridging,yun2019improvement,wang2017automatic,cui2019pulmonary,li2020model,zhang2021deep}. In these methods, three 2D CNNs are trained to segment from sagittal, coronal, and axial planes separately. After that, the segmentation results from each plane are fused to get the final segmentation results.
Instead of the most commonly used majority voting for ensemble learning, \cite{xia2018bridging} proposed Volumetric Fusion Net (VFN), a shallow 3D CNN with much fewer parameters for fusion of 2D results from different views. Their experimental results showed using VFN could obtain better 3D segmentation results than majority voting.

\begin{figure}
	\includegraphics[width=16cm]{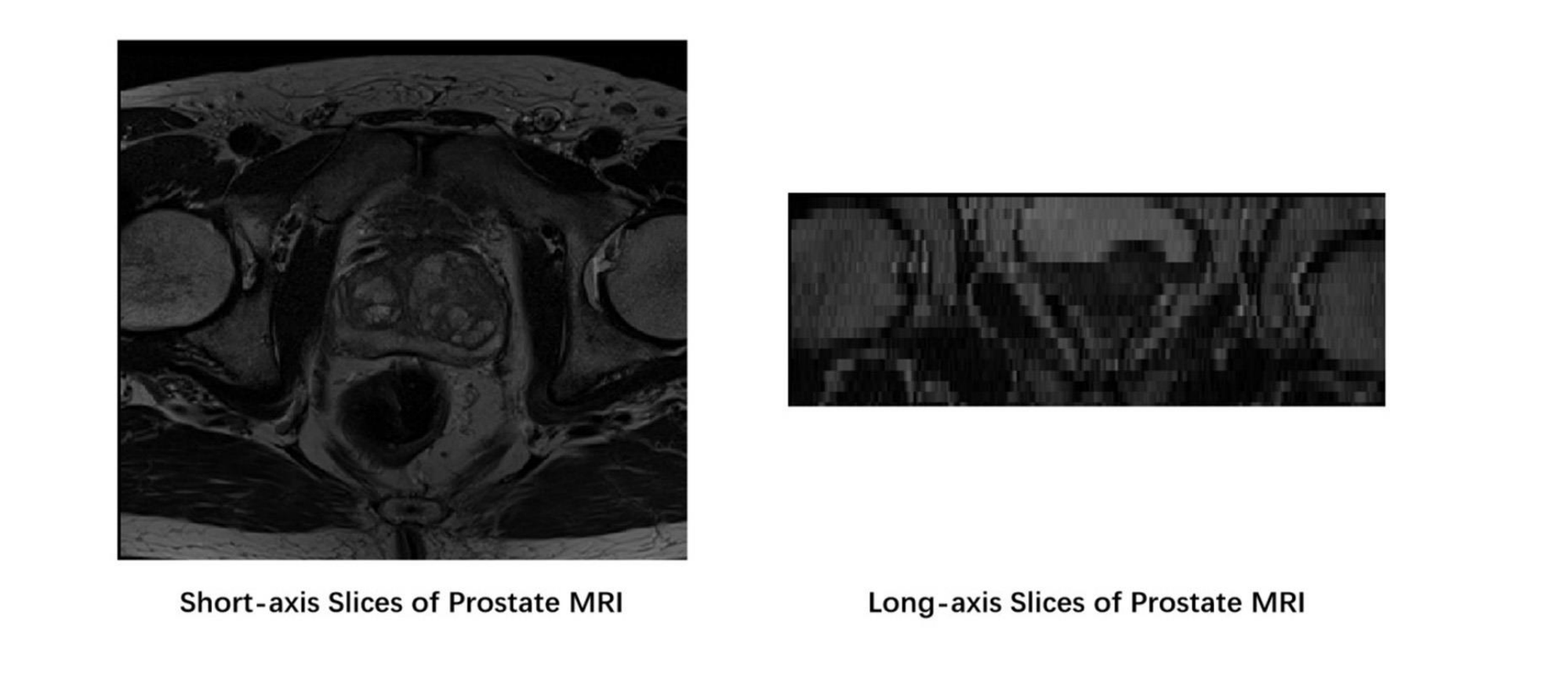}
	\centering
	\captionsetup{font={small}}
	\caption{Examples of short-axis slices and long axis slices for anisotropic volumetric medical images. From the figure we can see that long-axis slices are with poor contextual information due to the low resolution on z axis.}
	\label{Fig3}
\end{figure}

\subsection{Incorporating inter-slice information}

Due to the imaging settings, some modalities of medical images consist of anisotropic voxels, which means that the spatial resolution of three dimensions are not always the same. As a consequence, for these anisotropic volumetric images, the image resolution in x and y axis (length and width) is more than ten times higher than that of the z axis (depth), and x and y axis preserve much higher resolution and richer information than the z axis \cite{zhang2020cascaded}.
Therefore, neighboring regions along z axis may have abrupt changes around the same area \cite{ou2021lambdaunet}.
Figure \ref{Fig3} presents an example of short-axis and long-axis slices of prostate MR images with anisotropic voxels. From the figure, we can observe that long-axis slices suffer from poor contextual information due to the low resolution on z axis.
As a result, for anisotropic volumes, training 2D CNNs from long-axis slices (xz or yz planes) may be inadvisable.

\begin{figure}
	\includegraphics[width=16cm]{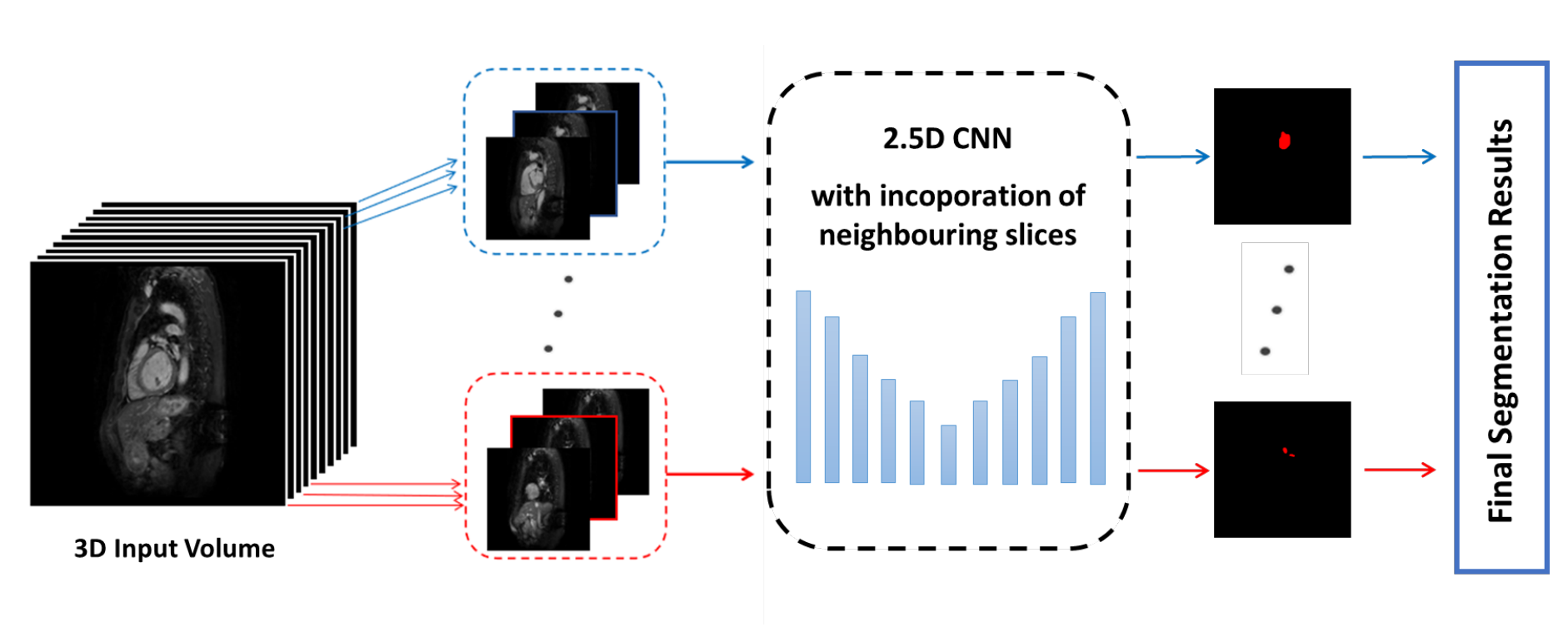}
	\centering
	\captionsetup{font={small}}
	\caption{The overall workflow of incorporating inter-slice information. With the additional input of neighboring slices, the inter-slice correlation could be leveraged so as to get better performance compared with 2D CNNs.}
	\label{Fig4}
\end{figure}

Instead of fusing results from different segmentation views, another strategy is incorporating inter-slice information into 2D CNNs to explore the spatial correlation. The workflow of this strategy is shown in Figure \ref{Fig4}.
In this way, the network could utilize not only the information of the slice at hand, but also the information of its neighboring slices. The inputs of the network are consecutive slices while the output is the corresponding segmentation results of the middle slice. By incorporating inter-slice information, the spatial correlation of the volume can be leveraged while avoiding the heavy burden of 3D computing.

The most widely used method draws on the idea of natural image segmentation and introduces sequential slices as multi-channel input for the segmentation of the middle slice. Using multi-slice input for utilization of spatial information has been applied in many medical segmentation tasks \cite{duan2019automatic,wang2019volumetric,yu2018recurrent,li2021ace,zhou2021intracranial}. Most of these methods feed 3 slices into segmentation networks: the slice at hand and its adjacent slices \cite{liu2019automatic,li2021learning}.
Instead of inputting continuous slices, Zhao \textit{et al.} \cite{zhao2021multi} proposed to utilize both dense sampling with adjacent slices and sparse sampling with dilated slices for 2.5D segmentation.
However, directly adding adjacent slices as multi-channel input may be inefficient. When neighboring slices are mixed together into channel dimension, the information of input slices are fused in the first convolution layer. This process makes the network harder to pick up useful information for distinguishing each slice, which may lead to worse performance. Therefore, some work has focus on designing new architectures for inter-slice information extraction.

The successful application of Recurrent Neural Networks (RNNs) in segmentation tasks \cite{yu2018recurrent,yang2018towards} provides potential solutions for refinement of 2D segmentation results. In this way, 2D slices of the 3D volume are viewed as a time series sequence to distill the inter-slice contexts and features. Among the variants of RNNs, bi-directional convolutional long short-term memory (BC-LSTM) \cite{chen2016combining} exhibits better improvement as the network can integrate information flow from two directions to improve the spatial continuity of the overall segmentation results, which has been applied in many medical image segmentation tasks \cite{zhu2018exploiting,li2020model}.

Other than using RNNs, another line of researches focuses on using attention mechanism to utilize inter-slice information. Since the information of each slice is spatially correlated with its upper and lower slices due to the spatial continuity, the information of adjacent slices could be used to guide the segmentation procedure by highlighting the most salient areas using attention mechanism. Zhang \textit{et al.} \cite{zhang2020saunet} proposed an inter-slice attention module that uses the information of adjacent slices to generate the attention masks so as to provide a priori shape regulation for the segmentation. Kuang \textit{et al.} \cite{kuang2020psi} applied contextual-attention block to force the model to focus on the border areas using element-wise subtraction between slices. Using attention mechanism has been an emerging trend for extraction of inter-slice information to obtain further improvements.

\begin{figure}
	\includegraphics[width=16cm]{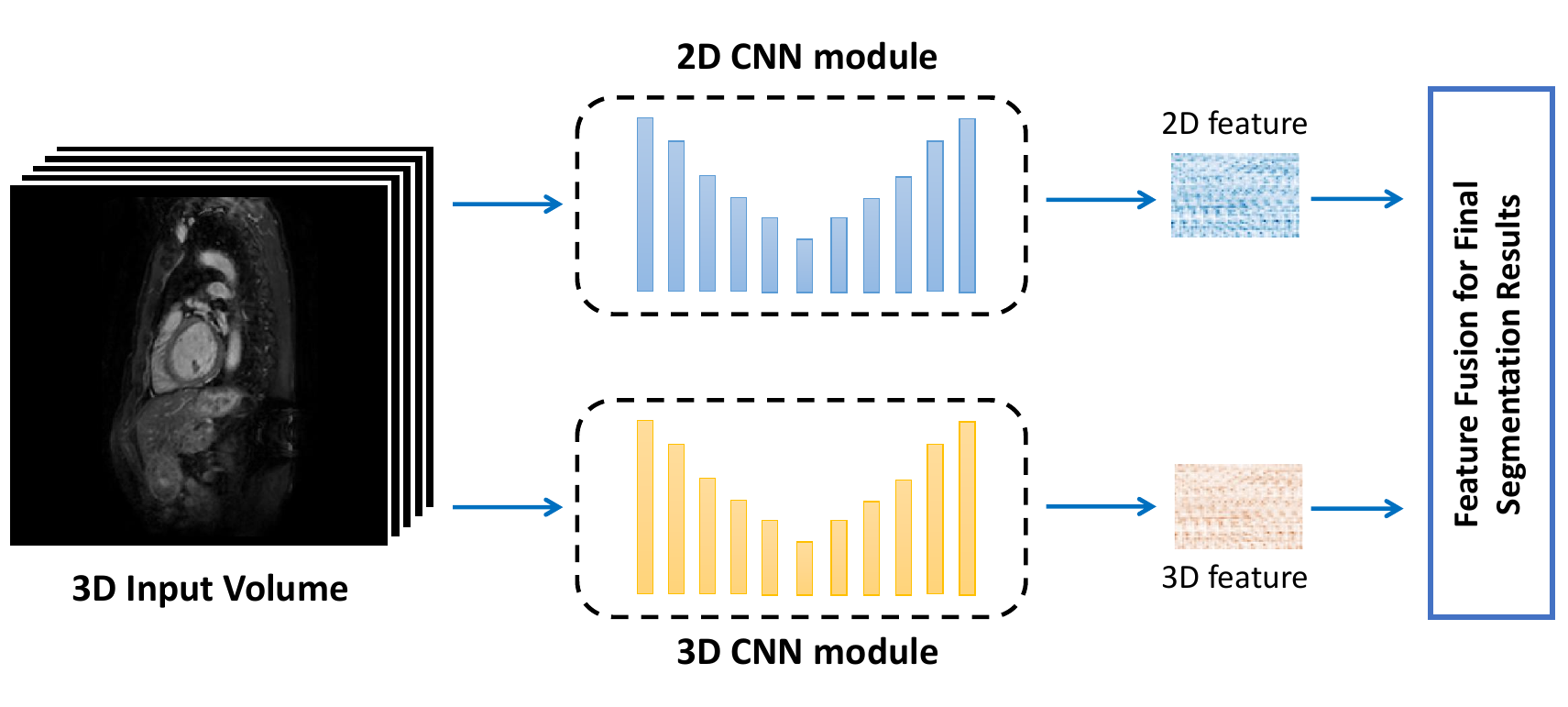}
	\centering
	\captionsetup{font={small}}
	\caption{The overall workflow of fusing 2D/3D features. Features extracted from 2D and 3D CNNs/modules are fused to utilize volumetric spatial information.}
	\label{Fig5}
\end{figure}

\subsection{Fusing 2D/3D features}

For computation-efficient volumetric medical image segmentation, some works also focus on the fusing features extracted from 2D and 3D CNNs to obtain higher efficiency \cite{li2018h-denseunet,zhou2019d-unet,chen2021novel,mei2021automatic}. Figure \ref{Fig5} presents the workflow of this strategy. Although these methods still use 3D convolutions to extract spatial information, however, the overall computation cost is reduced compared with training pure 3D CNNs. In this way, this strategy can be regarded as 2.5D methods to explore the segmentation efficiency.
The first method uses 2D results to provide priori shape information for 3D CNNs, then 3D CNNs are used to utilize the spatial information based on 2D outputs and the original volume. In the end, 2D and 3D results are fused to get the final segmentation results \cite{li2018h-denseunet}. Another way uses use both 2D and 3D encoder for the feature extraction. Then the 2D and 3D features at different scales are combined respectively in the encoding stage \cite{zhou2019d-unet}. Besides, instead of directly fusing the features, squeeze-and-excitation (SE) block \cite{hu2018squeeze} is applied to achieve better fusion by weighting the output of the two dimensions. In summary, both methods aim at combining 2D and 3D features for efficient extraction of volumetric information. The main difference is the stage and method to fuse the 2D and 3D features.

\section{Experiments}

Despite many 2.5D methods are proposed for efficient volumetric medical image segmentation, these methods are evaluated on different datasets with different settings and backbone structures. The comparison between these methods has not been verified. Therefore, in the following experiments, we intend to systemically evaluate the performance of these 2.5D methods with the same backbone structure and settings to compare their performance on three representative segmentation datasets.

\subsection{Datasets}

In the following experiments, we use 3 representative datasets of Medical Segmentation Decathlon Challenge \cite{simpson2019large} for the evaluation of these methods.
The first dataset is the Cardiac which includes 20 mono-modal MR volumes provided by King’s College London, for the segmentation of left atrium. The left atrium appendage, mitral plane, and portal vein end points were segmented by an expert using an automated tool \cite{ecabert2011segmentation} followed by manual correction.
The second dataset is Spleen which includes 41 CT volumes from Memorial Sloan Kettering Cancer Center for spleen segmentation. 
The spleen was semi-automatically segmented using the Scout application and the contours were manually adjusted by an expert abdominal radiologist.
The third dataset is Prostate which includes 32 Multimodal MR (T2, ADC) volumes provided by Radboud University for segmentation of prostate.
Specifically, the Prostate dataset has highly anisotropic voxel spacings with a typical image shape of 16x320x320.
All the datasets are randomly splitted at a rate of 80\% for training and 20\% for testing. We chose these three datasets because we want to involve typical modalities (CT and MRI) and targets (cardiac, prostate and abdomen) in 3D medical image segmentation tasks so as to obtain more general conclusions.

\subsection{Backbone structure and evaluation metrics}

We employ U-Net \cite{ronneberger2015u-net} as the network backbone that has four stage convolutional blocks in different resolutions and the base convolution block has 8 feature maps. For comparison of 3D CNNs, we employ 3D U-Net \cite{iek20163D} on the same experimental environment and settings.
All the experiments are implemented using Pytorch and run in Linux on NVIDIA Tesla V100 GPUs. During training, we use the Adam optimizer for all experiments with initial learning rate 0.001 and would be dropped by the factor of 0.5 if the training loss was no more improved in 20 epochs. We use the combination of cross-entropy loss and dice loss as the loss function for training. 
For quantitative evaluation of segmentation results, two complementary metrics are introduced. Dice coefficient (DSC) are used to measure the region mismatch and 95\% Hausdorff Distance (95HD) are used to evaluate the boundary errors between the segmentation results and the ground truth. 
Higher DSC (closer to 1) indicating more complete overlap and lower 95HD (closer to 0) indicating closer boundaries represent better segmentation performance.
The evaluation using these two metrics can achieve a more comprehensive comparison of segmentation results.

\subsection{Experimental design}

For the first experiment, we aim to compare the performance of 2D segmentation results from different segmentation planes including sagittal (S), coronal (C) and axial (A) planes, and the 2.5D multi-view fusion results with different fusion methods including majority voting (MV), weighted voting (WV) and Volumetric Fusion Net (VFN) \cite{xia2018bridging}.
For the second experiment, we first compare the performance of different number of input slices for multi-slice input. Then we compare the performance of different ways to incorporate inter-slice information including multi-slice input, applying RNNs and using attention mechanism as in \cite{zhang2020saunet} and \cite{kuang2020psi}.
Specifically, we select the segmentation plane with best performance according to the former experimental results to generate 2D slices.
After that, for the third experiment, we make comparisons of different stages including feature fusion at encoder stage and output stage, and different methods including Add and SE \cite{hu2018squeeze} to fuse 2D and 3D features.

\section{Results}

In this section, we systematically evaluate the performance of these 2.5D methods on three representative public datasets with the above implementation details.

\subsection{Experimental results of multi-view fusion}

\begin{table}[]
\captionsetup{font={normalsize}}
\caption{Quantitative segmentation results of 2D U-Net from different planes, 2.5D methods with different fusion strategies and 3D U-Net on the evaluation datasets. The arrows indicate which direction is better.}
\label{Table1}
~\\[1pt]
\centering
\setlength{\tabcolsep}{2.2mm}{
\renewcommand \arraystretch {1.2}
\begin{tabular}{ccccccccc}
\hline

\multirow{2}{*}{Dataset} & \multicolumn{1}{l}{\multirow{2}{*}{Metrics}} & \multicolumn{3}{c}{2D CNNs}                   & \multicolumn{3}{c}{2.5D Multi-view Fusion} & 3D CNNs  \\

    & \multicolumn{1}{l}{}  & U-Net (A)  & U-Net (S)  & U-Net (C) & MV   & WV  & VFN  & 3D U-Net \\ \hline
Cardiac MRI & $Dice \uparrow$ & 0.8362 & \textbf{0.8630} & 0.8225 & 0.8992 & 0.8996 & \textbf{0.9140} & 0.9203  \\
\textbf{}  & $95HD  \downarrow$  & 7.120  & \textbf{3.274}  & 5.268 & 1.287 & 1.104 & \textbf{1.021} & 1.014 \\ \hline
Spleen CT & $Dice \uparrow$  & \textbf{0.9114} & 0.8027 & 0.8858  & 0.9204 & 0.9208 & \textbf{0.9410} & 0.9452 \\
\textbf{}  & $95HD \downarrow$ & \textbf{1.521} & 8.710  & 2.165 & 1.511 & 1.508 & \underline{\textbf{1.119}} & 1.222 \\ \hline
Prostate MRI & $Dice \uparrow$  & \textbf{0.7222} & 0.4322 & 0.4545 & 0.6448 & 0.7490 & \underline{\textbf{0.7731}} & 0.7677 \\
\textbf{} & $95HD  \downarrow$ & \textbf{8.933} & 23.72 & 15.31 & 4.956 & 4.402 & \underline{\textbf{4.460}} & 5.935 \\ \hline
\end{tabular}}

\end{table}

Table \ref{Table1} presents the quantitative results of multi-view fusion. The experiments are designed to evaluate the performance of multi-view fusion on different segmentation tasks. Firstly, we train 2D U-Net from sagittal, coronal and axial planes, named U-Net(S), U-Net(C) and U-Net(A) respectively. Then 2.5D multi-view fusion results are obtained by ensembling results from three segmentation views using different fusion methods.
Compared with the best 2D results of different views, multi-view fusion with majority voting can obtain performance improvement by 3.62\% and 0.90\% in terms of Dice on Cardiac MRI and Spleen CT datasets, respectively.
Compared with majority voting, using VFN could achieve additional performance gains by 1.48\% and 2.06\% on Dice, although we need to train an extra network for fusion of different outputs, resulting in additional computation cost.
However, for anisotropic volumes like Prostate MRI, we observe that the performance of 2D CNNs for sagittal and coronal slices is poor due to limited intra-slice information. When taking the unsatisfactory results into account for multi-view fusion, the performance may be even worse compared with the best 2D results on Dice metric. To issue the problem, we use another fusion method by giving the short-axis slices (generally with better 2D results) more weight, which we named weighted voting (WV). From the results, we can observe that weighted voting outperforms majority voting, since the additional weight can mitigate the impact of bad results while fusing useful multi-view information. However, for Cardiac MRI and Spleen CT datasets, the improvement of WV compared with MV is not obvious.

\subsection{Experimental results of incorporating inter-slice information}

\begin{table}[]
\captionsetup{font={normalsize}}
\caption{Quantitative segmentation results of 2D and 2.5D methods with different number of input channels on the evaluation datasets. The arrows indicate which direction is better.}
\label{Table2}
~\\[1pt]
\centering
\setlength{\tabcolsep}{3mm}{
\renewcommand \arraystretch {1.2}
\begin{tabular}{ccccccc}
\hline
\multirow{2}{*}{Dataset} & \multicolumn{1}{l}{\multirow{2}{*}{Metrics}} & 2D CNNs         & \multicolumn{3}{c}{2.5D Multi-slice Input}   & 3D CNNs  \\
                         & \multicolumn{1}{l}{}  & U-Net (1slice)  & 3 slices        & 5 slices       & 7 slices   & 3D U-Net \\ \hline
Cardiac MRI              & $Dice \uparrow$ & 0.8630    & \textbf{0.8781} & 0.8642      & 0.867    & 0.9203   \\
                         & $95HD \downarrow $   & 3.274   & \textbf{1.559}  & 2.525    & 3.837   & 1.014    \\ \hline
Spleen CT                &$ Dice \uparrow$ & 0.9114 & \textbf{0.9121} & 0.9082   & 0.9089   & 0.9452   \\
                         & $95HD \downarrow$   & 1.521  & \textbf{1.345}  & 1.581     & 1.527    & 1.222    \\ \hline
Prostate MRI             & $Dice \uparrow$   & 0.7222 & \textbf{0.7460} & 0.7448   & 0.7257   & 0.7677   \\
                         & $95HD \downarrow$   & 8.933  & 5.565   & \underline{\textbf{4.759}} & 7.225    & 5.935   \\ \hline
\end{tabular}}
\end{table}

\begin{table}[]
	\captionsetup{font={normalsize}}
	\caption{Quantitative segmentation results of 2.5D methods to incorporate inter-slice information, 2D U-Net and 3D U-Net. The arrows indicate which direction is better.}
	\label{Table3}
	~\\[1pt]
	\centering
	\setlength{\tabcolsep}{2.2mm}{
		\renewcommand \arraystretch {1.2}
		\begin{tabular}{cccccccc}
			\hline
			\multirow{2}{*}{Dataset} & \multicolumn{1}{l}{\multirow{2}{*}{Metrics}} & 2D CNNs         & \multicolumn{4}{c}{2.5D Incorporate Inter-slice Information}        & 3D CNNs     \\
			& \multicolumn{1}{l}{}  & U-Net           & Multi-slice     & RNN            & Att Shape      & Att Border      & 3D U-Net   \\ \hline
			Cardiac MRI              & $Dice \uparrow $   & 0.8630    & 0.8781 & \textbf{0.8858}   & 0.8821   & 0.8747   & 0.9203   \\
			&  $95HD \downarrow$   & 3.274     & 1.559 & 2.155   & \textbf{1.494}  & 1.750  & 1.014   \\ \hline
			Spleen CT                & $Dice \uparrow$  & 0.9114 & 0.9121 & 0.9167   & \textbf{0.9233}    & 0.9155          & 0.9452  \\
			& $95HD \downarrow$  & 1.521  & 1.345  & 1.573  & \textbf{1.390}   & 1.794    &  1.222   \\ \hline
			Prostate MRI             & $Dice \uparrow$     & 0.7222 & 0.7460 & 0.7801   & 0.7804    & \underline{\textbf{0.7885}} & 0.7677      \\
			& $95HD \downarrow$   & 8.933  & 5.565  & 4.914 & \underline{\textbf{3.448}} & 3.907   & 5.935    \\ \hline
	\end{tabular}}
\end{table}

To evaluate the influence of different number of input slices on incorporating contextual information, we make comparison experiments of different number of input slices for multi-slice input.
We select short-axis slices with the best 2D performance as the segmentation plane for the following experiments.
As shown in Table \ref{Table2}, generally, using with neighboring slices for multi-slice input can obtain better results compared with 2D single-slice input.
These findings prove that the usage of neighboring slices could improves the segmentation accuracy.
Interestingly, we observe that additional slices did not improve the performance when the number of input slices reaches three. One possible explanation is that non-adjacent neighboring slices with less useful information may sometimes mislead the segmentation of middle slice. Besides, more input slices make the computational cost of the network higher. As a result, using three as the number of input slices may be the best practice.

Table \ref{Table3} shows the quantitative results of different methods to incorporate inter-slice information and their comparison with 2D U-Net and 3D U-Net. In addition to naive multi-slice inputs, we conduct experiments on three different methods of incorporating inter-slice information: applying RNNs (Bi-CLSTM) for information extraction, using inter-slice contextual attention for priori shape regulation \cite{zhang2020saunet} and border areas \cite{kuang2020psi} . From the results, we could see that compared with naive multi-slice input, all three methods can obtain extra performance gains due to more efficient extraction of inter-slice information. However, these methods have variable performance on different datasets and none of them outperforms the other two on all tasks. Specifically, we observe that for segmentation of anisotropic volumes like Prostate MRI, 3D CNNs may not be the best choice since the scarce inter-slice information will be further lost when downsampling on z axis. As a result, all three methods outperform 3D U-Net on all metrics.

\subsection{Experimental results of fusing 2D/3D features}

\begin{table}[]
\captionsetup{font={normalsize}}
\caption{Quantitative segmentation results of different stages and methods to fuse 2D and 3D features. The arrows indicate which direction is better.}
\label{Table4}
~\\[1pt]
\centering
\setlength{\tabcolsep}{2.2mm}{
\renewcommand \arraystretch {1.2}
\begin{tabular}{cccccccc}
\hline
\multirow{3}{*}{Dataset} & \multicolumn{1}{l}{\multirow{3}{*}{Metrics}} & \multicolumn{1}{l}{\multirow{3}{*}{2D CNNs}}  & \multicolumn{4}{c}{Fusion of 2D and 3D features}  & \multicolumn{1}{l}{\multirow{3}{*}{3D CNNs}}                                              \\
                         & \multicolumn{1}{l}{}  &   & Encoder  & Encoder & Output & Output &        \\
                         & \multicolumn{1}{l}{}  &   & Add & SE  & Add  & SE  &          \\ \hline
Cardiac MRI              & $Dice \uparrow$      & 0.8630 & 0.8743  & 0.8732  & \textbf{0.8876}  & 0.8824  & 0.9203        \\
                         & $95HD \downarrow$    & 3.274  & 1.759   & 1.913   & \textbf{1.604}   & 1.707   & 1.014       \\ \hline
Spleen CT                & $Dice \uparrow$      & 0.9114 & \textbf{0.9221}  & 0.9193  & 0.9197  & 0.9122  & 0.9452     \\
                         & $95HD \downarrow$    & 1.521  & 1.445   & 1.463   & \textbf{1.425}   & 1.487   & 1.222   \\ \hline
Prostate MRI             & $Dice \uparrow$      & 0.7222  & 0.7562  & \underline{\textbf{0.7693}}  & 0.7625  & 0.7597  & 0.7677  \\
                         & $95HD \downarrow$    & 8.933  & 7.141   & 6.259   & \underline{\textbf{5.567}}   & 8.780   & 5.935 \\ \hline
\end{tabular}}
\end{table}

In this subsection, we aim at comparing the performances of different stages and methods to fuse 2D and 3D features. For feature fusion stages, we select fusing features extracted from 2D and 3D encoders at encoding stage as in \cite{zhou2019d-unet} and fusing the output feature maps of 2D and 3D networks as in \cite{li2018h-denseunet}. For feature fusion methods, two methods of directly adding corresponding feature maps and using squeeze-and-excitation (SE) block \cite{hu2018squeeze} are compared. Table \ref{Table4} shows the evaluation results of fusing 2D and 3D features. From the table we could see that compared with 2D CNNs, all fusion methods can achieve better performance. Specifically, for Cardiac MRI segmentation, fusing features at output stage could get better results than that at encoding stage. However, for Spleen CT and Prostate MRI, these methods have variable performance and no one is superior to the others on all tasks. 
As a conclusion, fusing 2D and 3D features could efficiently utilize volumetric spatial information and improve the performance compared with 2D CNNs. For different fusion methods, fusing at encoding stage could urge the network to extract useful information with affordable computation cost. For fusing 2D and 3D outputs, the addition of 2D results could provide useful intra-slice information and accelerate the training procedure of 3D network. Although all these methods can improve the segmentation efficiency, however, the comparison of their performance is still affected by the characteristics of segmentation tasks.

\begin{table}[]
	\captionsetup{font={normalsize}}
	\caption{Comparison of model complexity of different segmentation methods. }
	\label{Table5}
	~\\[1pt]
	\centering
	\setlength{\tabcolsep}{3.2mm}{
		\renewcommand \arraystretch {1.2}
		\begin{tabular}{cccc}
			\hline
			Methods & CNNs & Parameters (M) &  Training Time \\ \hline
			2D U-Net  & 2D & 0.487 & 1.00x \\
			3D U-Net  & 3D & 1.403 & 4.14x \\ 
			VFN (shallow 3D U-Net) & 3D & 0.088 & ——  \\ \hline
			2D U-Net + Att Shape & 2.5D  & 0.487  & 1.11x \\
			2D U-Net + Att Border & 2.5D  & 0.487  & 1.08x \\
			2D U-Net + RNN & 2.5D  & 0.535 & 1.31x \\
			2D/3D Fusion (Encoder) & 2.5D  & 0.543 & 1.54x \\
			2D/3D Fusion (Output) & 2.5D & 0.841  & 2.32x \\ \hline
	\end{tabular}}
\end{table}

\section{Conclusion and Discussion}

In this paper, we aim at exploring 2.5D method for efficient volumetric medical image segmentation to overcome the lack of volumetric information of 2D CNNs and heavy computation cost of 3D CNNs, and make a large-scale empirical comparison on three representative medical image segmentation tasks involving different modalities and targets. Generally, 3D CNNs seem to be the best choice for segmentation accuracy, however, the unaffordable computational cost remains a hindrance for their further application. Besides, for anisotropic volumes that x and y axis preserve much higher resolution than the z axis, the performance advantages of 3D CNNs are not obvious, sometimes even worse than some 2.5D methods. For these anisotropic images with high discontinuity between slices, directly applying 3D CNNs may involve irrelevant features and hinder the learning process, which is in line with the conclusions in \cite{zhang2020saunet,ou2021lambdaunet}.

For 2.5D segmentation methods, we observe that all these 2.5D methods have the potential to improve the performance of 2D baseline. Multi-view fusion of 2D results from different views can get segmentation results comparable to 3D CNNs. However, we still need to train 3 (or 4 using VFN) networks to get the final segmentation results. In addition, the improvement of this strategy for anisotropic volumes is not significant.
For incorporating inter-slice information, we could obtain performance improvement while training only one network with slight computational increasement. Compared with naive multi-slice input to incorporate volumetric information, all RNN-based and attention-based 2.5D methods can further improve the segmentation performance. However, the performance gains of these methods are not consistent in different datasets.
For fusing 2D/3D features, all these fusion methods can improve the segmentation efficiency compared with using only 2D CNNs, demonstrating the effectiveness of fusing 3D features to utilize volumetric information. However, the performance of different fusion methods is affected by the characteristics of different segmentation tasks. Table \ref{Table5} presents the comparison of model parameters and average training time on different datasets (ratio to 2D baseline) of different segmentation methods. It can be observed that all 2.5D methods have fewer parameters and less computation cost compared with 3D CNNs.
We cannot claim that we have completely reproduced all these methods, however, we tried our best to tune each method so as to ensure a fair comparison. Another limitation is that we only use 2D/3D U-Net as the backbone structure. However, they may not always be the best choice for all segmentation tasks. Still, we hope the results and conclusions of our study will prove useful for the community on exploring and developing efficient volumetric medical image segmentation methods.

\section*{Acknowledgement}

This work is supported by the National Key Research and Development Program of China under Grant 2016YFF0201002, the University Synergy Innovation Program of Anhui Province under Grant GXXT-2019-044, and the National Natural Science Foundation of China under Grant 61301005.

\bibliography{mybibfile}

\end{document}